# СТРУКТУРА МАГНИТНЫХ НЕОДНОРОДНОСТЕЙ В ПЛЕНКАХ С ТОПОЛОГИЧЕСКИМИ ОСОБЕННОСТЯМИ

### Е.Б. Магадеев, Р.М. Вахитов

**Введение**

В настоящее время ведутся активные исследования структуры и свойств вихреподобных неоднородностей (магнитных вихрей, скирмионов, бимеронов и т.д.), возникающих в некоторых классах магнетиков, что связано с реальными перспективами их использования в различных устройствах спинтроники, а также магнитной памяти нового поколения [1]. Такой интерес обусловлен их топологической защищенностью, наноразмерами, высокой подвижностью и другими уникальными спин-электронными свойствами [2-4]. В данной работе, в продолжение идей, изложенных и рассмотренных в [5] на примере простейшей одномерной модели, предлагается еще один тип вихреподобных неоднородностей, образующихся в тонких магнитных пленках с искусственно созданными отверстиями (антидотами [6]) или немагнитными включениями.

**1. Континуальная модель**

Пусть материал тонкой пленки с одним отверстием представляет собой ферромагнетик с сильной одноосной анизотропией типа «легкая плоскость», благодаря чему вектор намагниченности почти не выходит из плоскости пленки. Тогда энергия магнетика может приближенно представлена в следующем виде [5]:

$$E = \int A(\nabla\theta)^2 h dS, \qquad (1)$$

где угол $\theta$ задает ориентацию вектора намагниченности на плоскости, $A$ – обменный параметр, $h$ – толщина пленки. Здесь предполагается, что размагничивающие поля в рассматриваемом материале значительно меньше обменного взаимодействия.

В случае пленки без топологических особенностей функционал (1) имеет единственный минимум $E = 0$, достигаемый при $\theta = \text{const}$, что соответствует однородному распределению намагниченности. Однако при наличии даже одного отверстия в пленке (осуществимость таких наноразмерных перфораций на практике подтверждается аналогичными экспериментами с графеном [7]) это утверждение становится неверным. Введем на плоскости полярную систему координат $(r, \varphi)$ и примем, что магнетик заполняет область $R_{in} \le r \le R_{ex}$, имея тем самым форму проколотого диска. В этом случае кроме однородного состояния появятся также локальные минимумы, представляющие собой нетривиальные решения уравнения Эйлера-Лагранжа для функционала (1), а именно – уравнения Лапласа $\Delta\theta = 0$. Дело в том, что при указанной топологии образца угол $\theta$ может уже и не быть однозначной функцией радиус-вектора. Вместо этого достаточно выполнения физически эквивалентного условия $\theta(r, \varphi + 2\pi) = \theta(r, \varphi) + 2\pi k$, где $k$ – целое число. Несложно видеть, что при заданном значении $k$, которое в данном случае имеет смысл топологического заряда, решением уравнения $\Delta\theta = 0$ является $\theta(r, \varphi) = k\varphi + \text{const}$.

Заметим, что решения, отвечающие разным значениям *k*, топологически неэквивалентны, и переходы магнетика между соответствующими состояниями были бы сопряжены с появлением разрывов у функции θ(*r*, φ) вдоль некоторой линии на диске. Тогда энергия образца в силу (1) стала бы бесконечно большой. Следовательно, в континуальном приближении все рассматриваемые состояния являются одинаково стабильными, независимо от значения *k*.

Энергия (1) магнитного диска в состоянии с топологическим зарядом *k* равна $E = 2\pi k^2 A h \ln(R_{ex}/R_{in})$, вследствие чего при увеличении $R_{ex}$ энергия образца будет неограниченно возрастать для всякого $k \neq 0$. Таким образом, изложенная теория может быть применена на практике исключительно к тонким кольцам с $R_{ex} \approx R_{in}$ (в предельном случае мы приходим к результатам, полученным в [5]), в то время как в неограниченных пленках с одним проколом физически реализуется только однородное состояние магнетика. Однако ситуация радикально меняется в том случае, когда пленка содержит не одно, а два отверстия. Пусть эти отверстия представляют собой цилиндрические антидоты радиуса $R_1$ и $R_2$, а их центры отстоят друг от друга на $a \gg R_1, R_2$ (рис. 1). Тогда в силу линейности уравнения $\Delta\theta = 0$ его решение может быть найдено как суперпозиция решений, полученных выше для случая пленки с одним антидотом: $\theta = k_1\varphi_1 + k_2\varphi_2$, где $\varphi_1, \varphi_2$ – полярные углы в системах координатах, связанных с центрами отверстий, а $k_1, k_2$ – целые топологические заряды. Рассмотрим асимптотическое поведение полученного решения на большом удалении от системы, когда $r_1 \approx r_2 = r \gg a$. При этом $\varphi_1 \approx \varphi_2 = \varphi$, так что $\theta = (k_1 + k_2)\varphi$, и в случае $k_1 + k_2 \neq 0$ энергия неограниченной пленки вновь оказывается бесконечной. Чтобы избежать этого, положим $k_1 = k$, $k_2 = -k$, тогда

$$\theta = k(\varphi_1 - \varphi_2). \tag{2}$$

Из рис. 1 ясно, что входящая в это соотношение разность углов равна углу, под которым из данной точки виден отрезок, соединяющий центры отверстий, так что $\theta \sim r^{-1}$, а значит, $(\nabla\theta)^2 \sim r^{-4}$, что обеспечивает сходимость интеграла (1) в области больших *r*. Таким образом, энергия магнетика в рассматриваемом состоянии является конечной. Детальный расчет, показывает, что она равна

$$E = 4\pi k^2 A h \ln\left(a/\sqrt{R_1 R_2}\right). \tag{3}$$

Распределение намагниченности в окрестности отверстий, задаваемое выражением (2) при $k = 1$, схематически показано на рис. 1. Состояние с $k = -1$ получается из него симметричным отражением; $k = 0$ отвечает однородно намагниченному образцу. Состояния с другими значениями *k* для дальнейшего изложения интереса не представляют.

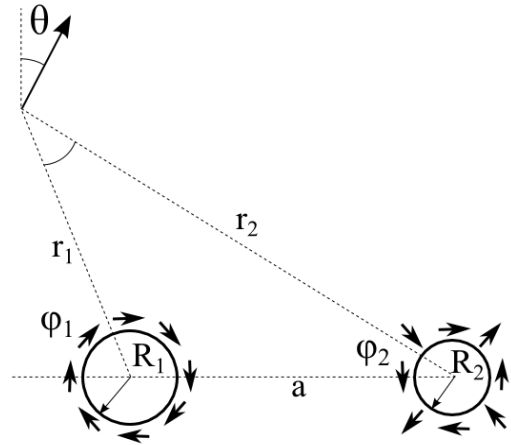

**Рис. 1.** Схема пленки с двумя отверстиями

## 2. Дискретная модель

Переходя к решеточной модели [8], будем рассматривать квадратную решетку спинов, некоторые узлы которой оставим пустыми. Тогда, с точностью до множителя, энергия магнетика за вычетом энергии однородного состояния будет иметь вид:

$$E = \sum_{i,j}\left(1 - \cos(\theta_i - \theta_j)\right) + \sum_i(1 - \cos\theta_i), \qquad (4)$$

где суммирование в первом слагаемом ведется по всем парам узлов решетки, соседствующим по горизонтали или вертикали, а во втором – по всем узлам, находящимся на краю решетки. Второе слагаемое, благодаря которому однородное состояние магнетика оказывается определенным однозначно ($\theta_i = 0$), позволяет имитировать бесконечную протяженность пленки: мы предполагаем, что неоднородности локализуются в области решетки, в то время как за ее пределами все спины ориентированы одинаково.

Будем оптимизировать энергию (4) численно. Поскольку нас интересуют все локальные минимумы системы, воспользуемся стохастическим подходом, при котором численная оптимизация повторяется многократно, начинаясь из различных стартовых точек, выбираемых случайно. Далее все найденные минимумы ранжируются по значениям энергии, и для каждого значения выбирается по одному решению (мы предполагаем, что существование неэквивалентных минимумов, равных по величине, крайне маловероятно). В результате такого поиска для решетки, содержащей 188 спинов, помимо однородного состояния с $E = 0$ обнаруживается также локальный минимум с $E = 10.3$. Соответствующее ему распределение намагниченности показано на рис. 2. Несложно видеть, что оно в точности воспроизводит состояние с $k = 1$, предсказанное нами в рамках континуальной модели.

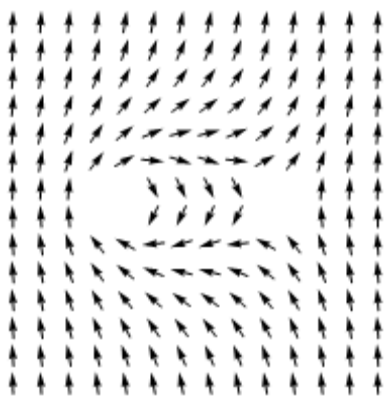

**Рис. 2.** Уединенная неоднородность, локализованная на двух отверстиях

Несмотря на то, что дискретная модель дает независимое подтверждение развиваемой нами теории, ее основное достоинство заключается не в этом. Значение выражения (4) ограничено не только снизу, но также и сверху, благодаря чему переходы между любыми состояниями решетки могут сопровождаться совершением лишь конечной работы. Это позволяет сделать оценки устойчивости неоднородных состояний, которые считаются метастабильными; такой подход, использованный ранее в [5], позволил получить нетривиальные результаты для случая замкнутой цепочки спинов (одномерная модель). В наиболее грубом приближении мы можем рассмотреть преобразование $\theta_i \to (1 - t)\, \theta_i$, которое является непрерывным и связывает между собой неоднородное состояние при $t = 0$ с однородным при $t = 1$. Тогда $E$ становится функцией параметра $t$ и, вследствие минимальности на обоих концах отрезка $[0; 1]$, должна иметь на нем также и максимум. В нашем случае значение этого максимума оказывается равно $E_{max} = 11.6$, поэтому работа, которую необходимо совершить, чтобы разрушить неоднородное состояние, составляет около $E_{max} - E = 1.3$. Подбирая более сложные преобразования с использованием численных методов, эту оценку можно улучшить до 1,0, однако в любом случае это значение составляет около 10% энергии самой неоднородности, что может говорить о неплохой устойчивости изучаемых состояний по отношению к тепловым флуктуациям.

**3. Влияние токов**

Покажем, что локализованные на двух отверстиях неоднородности с $k = \pm 1$ могут быть стабильными при воздействии тока, текущего через одно из отверстий. Пусть для начала сила такого тока чрезвычайно велика, так что влиянием обменного взаимодействия можно пренебречь. Тогда во всех точках образца вектор намагниченности ориентируется вдоль силовых линий магнитного поля, создаваемого током, то есть по касательным к окружностям, центры которых совпадают с центром отверстия. Тем самым это отверстие будет характеризоваться топологическим зарядом +1, а второе – зарядом 0, поскольку в его окрестности намагниченность будет почти постоянной. При уменьшении силы тока, однако, все большую роль будет играть обменное взаимодействие, энергия которого, как мы знаем, существенно понижается при суммарном топологическом заряде системы, равном 0. Ток, однако, все еще может оставаться достаточно сильным, чтобы вблизи него намагниченность по-прежнему ориентировалась вдоль силовых линий, так что топологический заряд, соответствующий первому отверстию, сохранит свое значение +1. Следовательно, для второго отверстия топологический заряд вынужденно станет равным –1 за счет спонтанного перераспределения намагниченности. Дальнейшее снижение силы тока приведет к тому, что вихрь намагниченности вокруг первого отверстия окажется невыгодным, и топологические заряды станут равными 0 для обоих отверстий; магнетик придет в состояние, близкое к однородному. Таким образом, мы ожидаем, что состояние с зарядами +1, –1 стабилизируется при значениях силы тока, лежащих в каком-то определенном интервале. При таких же значениях стабилизируется и состояние с зарядами –1, +1, однако ток следует пропускать уже через второе отверстие.

Проверим наши рассуждения посредством численного эксперимента. Для этого добавим в выражение (4) слагаемое следующего вида:

$$E_I = -\sum_i \frac{I}{r_i} \cos(\theta_i - \varphi_i),$$

где $(r_i, \varphi_i)$ – полярные координаты узлов в системе, связанной с точкой прохождения тока через плоскость, которую выберем в середине одной из пустот решетки. Оптимизируя полученную энергию при различных значениях тока $I$, мы обнаруживаем три неэквивалентных распределения намагниченности, которые показаны на рис. 3. В данном случае все распределения отвечают глобальному минимуму энергии, так что при указанных $I$ эти состояния являются стабильными. При других величинах силы тока, однако, состояния аналогичной топологии могут оказаться уже метастабильными. Зависимость энергии $E$ каждого из них от тока $I$ показана на рис. 4. Как и следовала ожидать при $I < 0.5$ устойчивым оказывается состояние с зарядами 0, 0; при $I > 2.6$ – с зарядами +1, 0; а при промежуточных значениях $0.5 < I < 2.6$ стабилизируется интересующая нас уединенная неоднородность, характеризующаяся зарядами +1, –1.

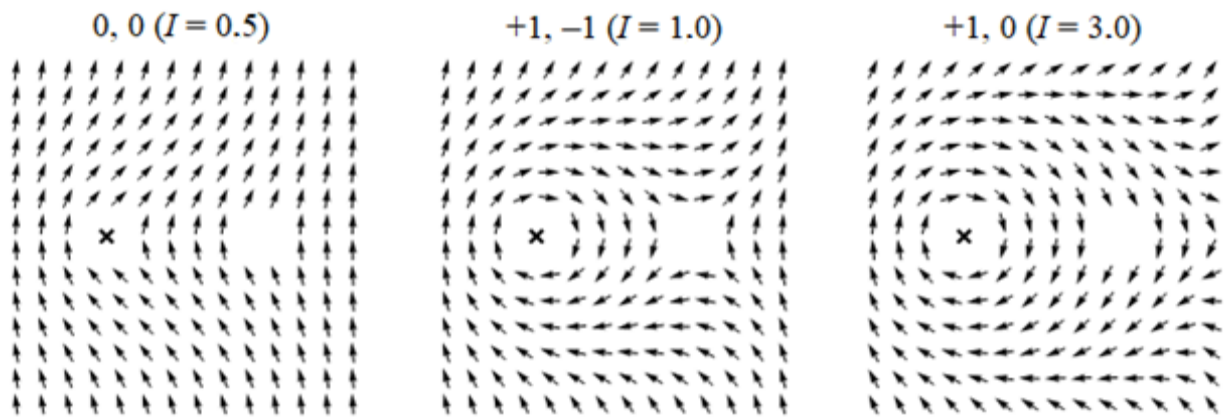

**Рис. 3.** Распределения намагниченности при воздействии тока

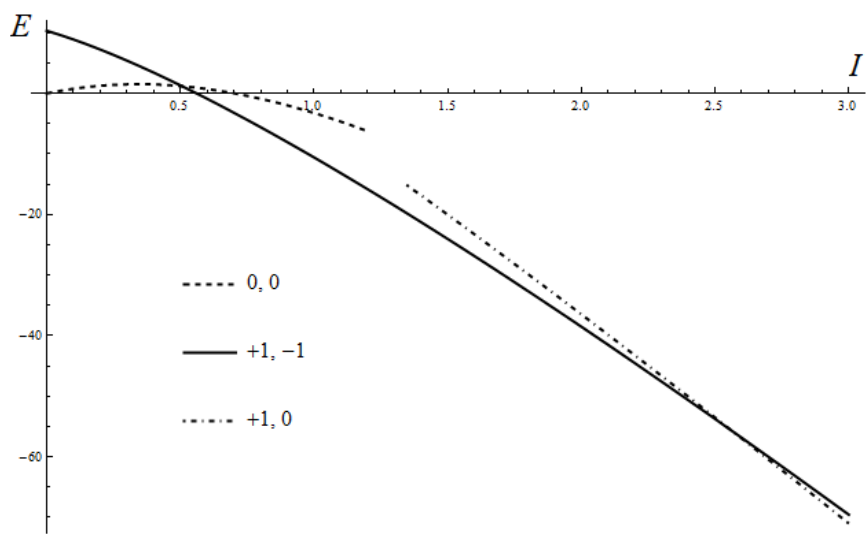

**Рис. 4.** Энергии различных состояний при воздействии тока

## 4. Случай четырех отверстий

Ясно, что все приведенные построения могут быть выполнены в отношении пленки не только с двумя, но также и с большим числом $N$ антидотов. При этом соотношение, аналогичное (2), будет иметь вид $\theta = k_1\varphi_1 + k_2\varphi_2 + \ldots + k_N\varphi_N$, вследствие чего должно выполняться равенство $k_1 + k_2 + \ldots + k_N = 0$, обеспечивающее конечность энергии системы.

Для примера рассмотрим случай $N = 4$, поместив отверстия равного радиуса $R$ в вершины квадрата со стороной $a$. Оптимизируя энергию (4) для соответствующей решетки из 180 спинов, мы получаем распределения намагниченности, показанные на рис. 5 (знаками отмечены отверстия, с которыми связаны заряды $\pm 1$). Энергии этих состояний равны $E_1 = 10.6$ (связаны два отверстия по горизонтали или вертикали), $E_2 = 13.5$ (связаны два отверстия по диагонали) и $E_3 = 15.9$ (связаны между собой все четыре отверстия).

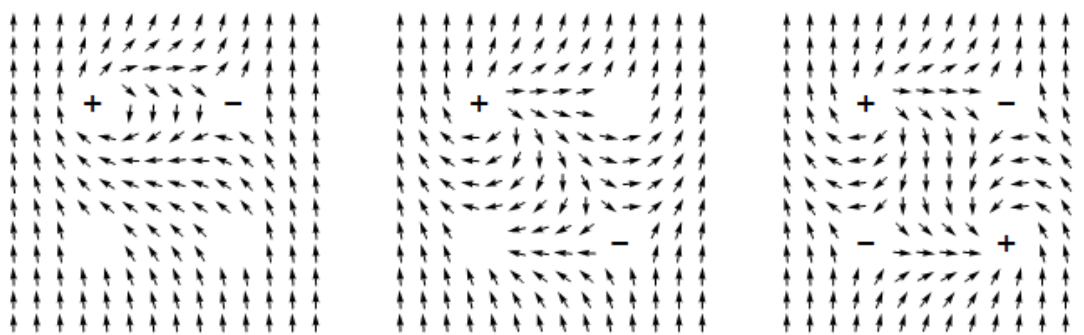

**Рис. 5.** Уединенные неоднородности, локализованные на четырех отверстиях

С точки зрения континуальной модели первые два из полученных состояний описываются соотношением (2) при $k = 1$, а значит, их энергии могут быть вычислены по формуле (3). Учитывая, что расстояние между центрами отверстий для второго состояния равно $a\sqrt{2}$, получаем: $E_1 = 4\pi A h \ln(a/R)$, $E_2 = 4\pi A h \ln(a\sqrt{2}/R)$. Для третьего состояния, можно показать, что $E_3 = 8\pi A h \ln(a/R\sqrt{2})$. Несложно видеть, что эта энергия может быть выражена через первые две следующим образом: $E_3 = 4E_1 - 2E_2$. Данное соотношение можно использовать в качестве контрольного для проверки соответствия между континуальной и дискретной моделями. Подставляя в него ранее найденные величины $E_1 = 10.6$ и $E_2 = 13.5$, имеем $E_3 = 15.4$, что лишь на 3% отличается от расчетного значения.

Несмотря на то, что из приведенных на рис. 5 неоднородностей третья обладает наибольшей энергией, именно она достойна пристального внимания. Дело в том, что эта структура, фактически, представляет собой связанное состояние двух неоднородностей первого типа, что подтверждается наличием энергии связи $E_3 - 2E_1 = -4\pi A h \ln 2 < 0$, а значит, разрушение такого состояния особо осложнено. Чтобы убедиться в этом, снова обратимся к минимальной работе, необходимой для перевода дискретной решетки спинов из данного состояния в однородное. «Грубая» оценка с использованием преобразования $\theta_i \to (1-t)\,\theta_i$ дает в этом случае сильно завышенное значение 12.5, которое удается улучшить до 4.0, что составляет примерно 25% энергии неоднородности и вчетверо превышает аналогичную работу для случая двух отверстий. Тем самым связывание сразу четырех отверстий, по всей видимости, легче осуществить на практике из-за большей относительной устойчивости возникающей неоднородности. При этом ее формирование, как явствует из рис. 5 и проведенного выше анализа влияния токов, может быть осуществлено пропусканием противоположно направленных токов определенной силы через два прокола, лежащих на одной диагонали квадрата.

**Заключение**

Таким образом, в работе показано, что в тонких пленках с топологическими особенностями в виде нескольких отверстий могут формироваться уединенные магнитные неоднородности, стабильные в присутствии тока и метастабильные в его отсутствие. При этом, однако, даже метастабильные состояния могут оказаться достаточно долгоживущими из-за существенного энергетического барьера, который необходимо преодолеть для разрушения неоднородности. Более того, этот барьер становится конечным только в рамках дискретной модели, выбор конкретных параметров которой остается в известной степени произвольным. По этой причине на данный момент не представляется возможным дать какие-то объективные оценки температур, при которых описанные состояния могут наблюдаться экспериментально. Именно этот теоретический вопрос, на наш взгляд, в наибольшей степени требует дальнейшей проработки, поскольку идея стабилизации неоднородных структур за счет факторов топологического характера пронизывает не только нашу работу, но и все построения, связанные со скирмионами. В то же время апеллировать к топологии непозволительно, когда речь идет о физических системах принципиально дискретной природы, каковыми отчасти являются и кристаллические решетки. Наличие последовательной методологии оценки надежности ограничений, накладываемых топологией на поведение магнитных структур, существенно упростило бы переход от теоретических предсказаний к их непосредственной экспериментальной проверке.

Тем не менее, есть определенная уверенность в отношении реализуемости на практике описанных нами неоднородностей. Их важной с технической точки зрения чертой является то, что они локализуются не в трехмерном пространстве, а на плоскости, благодаря чему становится реальным их использование в гетероструктурах с максимально тонкими слоями. Кроме того, каждая пара отверстий может кодировать более чем один бит, так как магнетик в их окрестности находится либо в однородном состоянии, либо в одном из двух неоднородных, отличающихся знаком топологического заряда. Тем самым вполне вероятно, что именно использование пленок с топологическими особенностями является ключом к повышению плотности записи информации, а также решению ряда других задач, связанных с развитием наноэлектроники.